\documentclass{osa-article}
\usepackage{bm}

\journal{optica}
\articletype{Research Article}

\begin{document}

\title{Controlling electron motion with attosecond precision by shaped
femtosecond intense laser pulse}
\author{Xiaoyun Zhao\authormark{1,\dag}, Mingqing Liu\authormark{2,\dag}%
,Yizhang Yang\authormark{3}, Zhou Chen\authormark{3}, Xiaolei Hao%
\authormark{1,6}, Chuncheng Wang\authormark{3,7}, Weidong Li\authormark{4},
Jing Chen\authormark{4,5,8}}

\address{\authormark{1}Institute of Theoretical Physics and Department of Physics, State Key
Laboratory of Quantum Optics and Quantum Optics Devices, Collaborative
Innovation Center of Extreme Optics, Shanxi University, Taiyuan 030006, China\\
\authormark{2}School of Physics and Information Technology, Shaanxi Normal University, Xi'an 710119, China\\
\authormark{3}Institute of Atomic and Molecular Physics and Jilin Provincial Key Laboratory
of Applied Atomic and Molecular Spectroscopy, Jilin University, Changchun
130012, China\\
\authormark{4}Shenzhen Key Laboratory of University Laser and Advanced Material Technology,
Center for Advanced Material Diagnostic Technology, and College of Engineering
Physics,Shenzhen Technology University, Shenzhen 518118, China\\
\authormark{5}Hefei National Laboratory,  Department of Modern Physics, University of Science and Technology of China, Hefei 230026, China\\
\authormark{\dag}These authors contribute equally to this work.\\
\authormark{6}xlhao@sxu.edu.cn\\
\authormark{7}ccwang@jlu.edu.cn\\
\authormark{8}chenjing@ustc.edu.cn} \medskip

%\email{\authormark{*}opex@osa.org} %% email address is required

%%%%%%%%%%%%%%%%%%%%%%%%%%  body  %%%%%%%%%%%%%%%%%%%%%%%%%%
\begin{abstract}
We propose the scheme of temporal double-slit interferometer to precisely
measure the electric field of shaped intense femtosecond laser pulse
directly, and apply it to control the electron tunneling wave packets in
attosecond precision. By manipulating the spectra phase of the input
femtosecond pulse in frequency domain, one single pulse is split into two
sub-pulses whose waveform can be precisely controlled by adjusting the
spectra phase. When the shaped pulse interacts with atoms, the two
sub-pulses are analogous to the Young's double-slit in time domain. The
interference pattern in the photoelectron momentum distribution can be used
to precisely retrieve the peak electric field and the time delay between two
sub-pulses. Based on the precise characterization of the shaped pulse, we
demonstrate that the sub-cycle dynamics of electron can be controlled with
attosecond precision. The above scheme is proved to be feasible by both
quantum-trajectory Monte Carlo simulations and numerical solutions of
three-dimensional time-dependent Schr\"{o}dinger equation.
\end{abstract}

\section{INTRODUCTION}

Precise control of electron motion in atoms or molecules became accessible
since the advent of laser. Considering the time scale of electron motion,
femtosecond or even attosecond lasers are indispensable to modulate electron
dynamics \cite{Niikura2002,Goulielmakis2010,Baltuska2003,Leone2014}. One can
directly change the parameter of a single laser \cite%
{Paulus2003,M2006,Kang2018,Blaga2009}, or construct a pump-probe scheme by
varying the time delay between two femtosecond lasers \cite%
{Zew1988,Zewail2000,Dantus1990} or one femtosecond laser plus one attosecond
laser \cite%
{Huh2010,Cavalieri2007,Kluender2011,Ossiander2016,Swoboda2010,Gong2022}.
Recently, direct combining two-color laser fields to form a shaped laser
field in temporal domain \cite%
{Mancuso2015,Lin2017,Eckart2018,Han2019,Milosevic2000}, which can be
controlled by varying the parameters of the two lasers, is applied to
modulate the motion of tunneling electron wave packets. On the other hand,
the technology of femtosecond pulse shaping \cite{Weiner1995,Weiner2000}, in
which Fourier synthesis methods are used to generate nearly arbitrarily
shaped ultrafast optical wave forms with a single laser, has been
successfully developed to manipulate the evolution of a quantum system \cite%
{Brixner2004,Dudovich2004} and thus control various atomic and molecular
processes such as the high harmonic generation \cite{Bartels2000,Oron2006},
the strong-field ionization of molecules \cite{Tagliamonti2017,Kaufman2020}.

How precise the control of electron dynamics can be achieved not only
depends on how subtly we can change the laser field but also depends on how
accurately we can characterize it. Since the electron dynamics is sensitive
to the temporal shape of the laser field \cite{Sola2006,Lin2017}, to control
the electron dynamics in attosecond precision puts higher requirements to
the control and measurement of the laser field in time domain. Femtosecond
pulse shaping is an ideal candidate due to its capability of temporally
reshaping the laser field in a very precise and flexible manner. However, to
direct measure the tailored laser field is still a challenge, which prevents
the further control of electron motion in a precise way. In practice, the
shaped laser field is reconstructed based on the information of the input
laser which are equally difficult to be measured accurately enough. As a
result, the shaped laser field can only be estimated in a largely uncertain
way which is far from the requirements. In this work, by introducing the
temporal double-slit interferometer, we propose a scheme to directly measure
the shaped laser field in an unprecedented precision. Based on this, control
of the electron tunneling wave packets in attosecond resolution with shaped
femtosecond laser pulse is achieved.

\section{Results and discussions}

\begin{figure}[tb]
\centering
\includegraphics[scale=0.9]{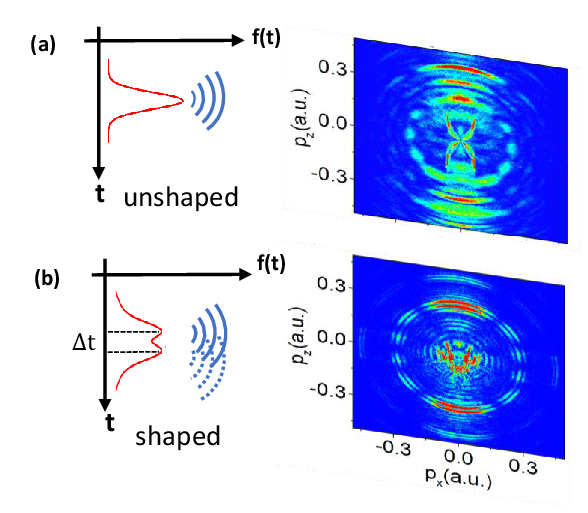}
\caption{Sketch maps for ionization processes if Xe atoms are
exposed to the unshaped single pulse (a) and to the shaped pulse
with the time delay of $ \Delta t$ between the two sub-pulses (b).
The photoelectron momentum distributions (PMDs) are simulated by
quantum trajectory Monte Carlo (QTMC) method. The laser intensity
of the input unshaped pulse is $ I_{0}=1.5\times10^{14}$
W/cm$^{2}$, the central wavelength is $\protect \lambda _{0}=800$
nm, and the pulse duration is 20 fs.} \label{fig1}
\end{figure}

As illustrated in Fig.~\ref{fig1}, a conventional single femtosecond intense
laser pulse can be shaped into a pair of sub-pulses by adding a phase of $\pi
$ in the spectra from specific wavelength $\lambda_{s}$ in frequency domain.
The shape of the output pulse, which is mainly determined by the amplitude
of the two sub-pulses and the time delay between them, can be controlled by
adjusting $\lambda_{s}$. The shaped laser field corresponding to specific $%
\lambda_{s}$ is measured base on the principle of Young's double-slit
interference in time domain. The basic idea behind it is as follows. If
atoms are exposed to unshaped single intense laser pulse, they will be
ionized and exhibit a typical photoelectron momentum distribution (PMD)
characterized by a series of above-threshold ionization (ATI) rings
separated by one photon energy \cite{M2006} as shown in Fig.~\ref{fig1}(a).
Meanwhile, the ATI ring displays stripes at specific angles, known as
jet-like structure. When the shaped pulse is used to ionize atoms, the
photoelectrons born in the two sub-pulses will interference, resulting in
the ATI ring to split into several sub-rings [see Fig.~\ref{fig1}(b)], while
the jet-like structure survives. The amplitude and the time delay of the two
sub-pulses can be retrieved based on positions and fringe spacing of the
sub-rings, just like what is done in Young's double-slit experiment for
light. We call this technique temporal Young's double-slit interferometer.
Then, by varying $\lambda_{s}$, very subtle changes of the shape of the
tailored pulse can be characterized. With this capability, we can manipulate
the electron tunneling dynamics in attosecond precision.

\begin{figure}[tb]
\centering
\includegraphics[scale=0.45]{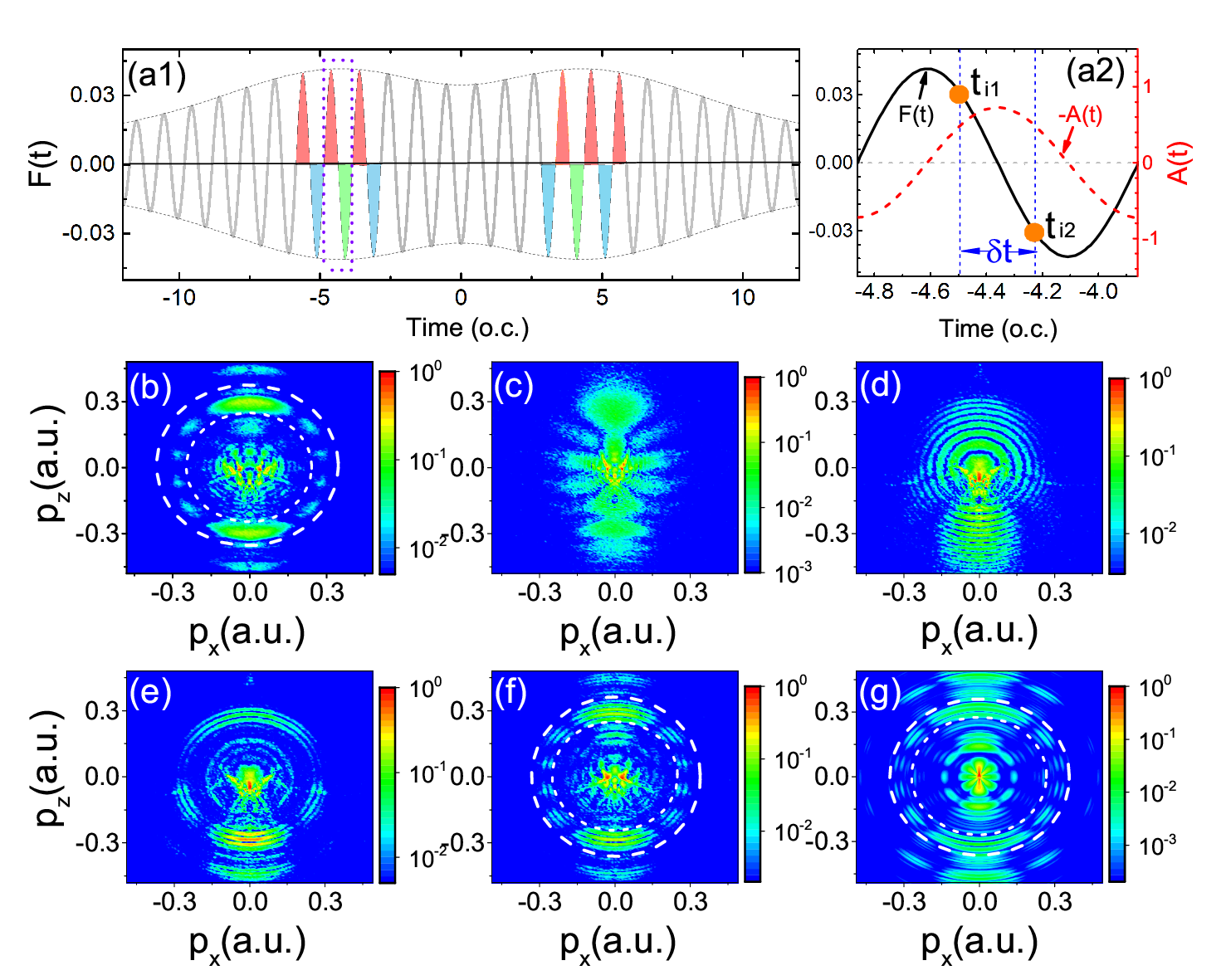}
\caption{(a1): The electric field $F(t)$ of the shaped pulse as a
function of time. (a2): The electric field and the negative vector potential
$-A(t)$ within one optical cycle indicated by violet dashed rectangle as
shown in panel (a1), $t_{i1}$ and $t_{i2}$ denote the ionization times of
intracycle interference trajectories which have the same final momentum with
time difference $\protect\delta t$. Simulated PMDs by QTMC are obtained by considering electrons
ionized at different time ranges: (b) $t_{i}<0$, (c) $t_{i}$ in a single
optical cycle enclosed by the dashed rectangle, (d) $t_{i}$ in green regions, (e) $%
t_{i}$ in green and blue regions, (f) $t_{i}$ in green, blue, and red
regions. (g): PMD calculated by numerically solving
three-dimensional time-dependent Schr\"{o}dinger equation (TDSE). The white
dashed lines are used to indicate the sub-ring and jet-like structure.
$\protect\lambda_{s}=780$ nm and the other pulse parameters are the same as
in Fig.~\protect\ref{fig1}.}
\label{fig2}
\end{figure}

Before that, we shall first demonstrate that the sub-ring
structure in Fig. \ref{fig1}(b) calculated with quantum trajectory
Monte Carlo (QTMC) method is indeed the result of the interference
of ionization events from different sub-pulses. In simulations of
QTMC \cite{li2014,song2016} (see supplement 1 for details), we can
explicitly obtain the PMD corresponding to ionization events born
in selected range of time in the shaped pulse. If the ionization
is confined to a single sub-pulse [for example, $t_{i}<0$ in
Fig.~\ref{fig2}(a1)], the PMD shows similar features to the case
of the unshaped pulse: clear ATI rings without splitting are
modulated with a jet-like structure [within white dashed lines in
Fig.~\ref{fig2}(b)]. The ATI rings with spacing of one photon
energy are usually recognized as the result of interference
between electrons ionized in adjacent optical cycles, termed as
intercycle interference \cite{Arbo2010}. The jet-like structure
can be attributed to the interference of electrons ionized in an
optical cycle \cite{li2014}, known as intracycle interference
\cite{Lindner2005}. As sketched in Fig.~\ref{fig2}(a2), electrons
emitted at times $t_{i1}$ and $t_{i2}$ possess an identical final
momentum of $ \mathbf{p}=-\mathbf{A}\left( t_{i}\right) $, where
$\mathbf{A}\left( t\right) $ is the vector potential, and
subsequently, they will interfere. This can be demonstrated
explicitly by considering electrons ionized within only one optical
cycle [the region in dashed box in Fig.~\ref{fig2}(a1)], as shown
in Fig.~\ref{fig2}(c), in which intracycle interference causes
the jet-like structure to prevail in the PMD.

Figure \ref{fig2}(b) suggests that ionizations in both
sub-pulses are necessary for the formation of the sub-ring
structure. Then, if the ionization times are limited in two
symmetric half-optical-cycles in the negative direction around the
envelope maximum of each sub-pulse [green regions in
Fig.~\ref{fig2}(a1)], the interference of electrons ionized in
these two regions generates a multitude of rings as shown in
Fig.~\ref{fig2}(d). The spacing of these rings matches that of the
sub-ring structure in Fig.~\ref{fig1}(b). Further, when electrons
originating from adjacent half-optical-cycles in negative field
direction [blue regions in Fig.~\ref{fig2}(a1)] are also involved,
adjacent intercycle interference comes into play and the sub-ring
structure is modulated by ATI rings [see Fig.~\ref{fig2}(e)].
Finally, after adding electrons ionized in the positive
electric field [red regions in Fig.~\ref{fig2}(a1)], intracycle
interference occurs and the jet-like structure [within white
dashed lines in Fig.~\ref{fig2}(f)] is formed. This almost
reproduces the complete PMD for the shaped pulse in
Fig.~\ref{fig1}(b). It can therefore be concluded that the
sub-ring structure in the PMD of the shaped pulse is induced by
interference between the electrons ionized in the two sub-pulses,
that is, double-slit interference in the time domain. Moreover, in
Fig.~\ref {fig2}(g), we also present simulated PMD by numerically
solving three-dimensional time-dependent Schr\"{o}dinger equation
(TDSE) \cite {jiang2017,liu2023} (see supplement 1 for
details of the method). The same laser parameters as that in QTMC
calculations are applied. The TDSE-simulated PMD, which also
features split ATI rings along with a jet-like structure [within
white dashed lines in Fig.~\ref{fig2}(g)], exhibits excellent
agreement with the QTMC results presented in Fig.~\ref {fig2}(g)
and Fig.~\ref{fig1}(b).

Based on the principle of Young's double-slit interference, the interference
fringes in the PMD are used to retrieve the peak field and time delay of the
shaped pulse constituted of two sub-pulses. The interference fringe spacing
is straitly related to the time delay between the two sub-pulses. The
relation can be obtained based on strong field approximation (SFA) \cite%
{Amini2019}. For electron ionized at time $t_{i}$ with final momentum $%
\mathbf{p}$, the phase is given by the classical action in atomic units ($%
\hbar =m_{e}=e=1$)
\begin{equation}
S(\mathbf{p},t_{i})=\int_{t_{i}}^{\infty }dt\left( \frac{1}{2}[\mathbf{p}+\mathbf{A}(t)]^{2}+I_{p}\right),  \label{act}
\end{equation}%
where $\mathbf{A}(t)$ is the vector potential, $I_{p}$ is the binding
potential. The interference fringe occurs when the phase difference between
electrons ionized around the two envelope maxima of the shaped pulse reaches
$2m\pi $ ($m$ is an integer), that is (see supplement 1 for details)
\begin{equation}
(E+I_{p}+U_{p})\Delta t-\frac{U_{p}}{\omega _{0}}\sin (\omega _{0}\Delta
t)=2m\pi ,  \label{fringe}
\end{equation}%
where $E=p^{2}/2$ is the electron energy, $U_{p}=F_{0}^{2}/4\omega _{0}^{2}$
is the pondermotive potential ($F_{0}$ is the peak electric field of the
sub-pulse and $\omega _{0}$ is the optical central frequency), $\Delta t$ is
the difference of the ionization time which is equal to the time delay
between the two sub-pulses. When the electron energy varies $\Delta E$ which
is just equal to the fringe spacing of the sub-ring structure in the PMD,
the phase difference will change $2\pi$. Then the time delay between the two
sub-pulses can be estimated as
\begin{equation}
\Delta t=2\pi /\Delta E.  \label{delay}
\end{equation}

\begin{figure}[tb]
\centering
\includegraphics[scale=0.44]{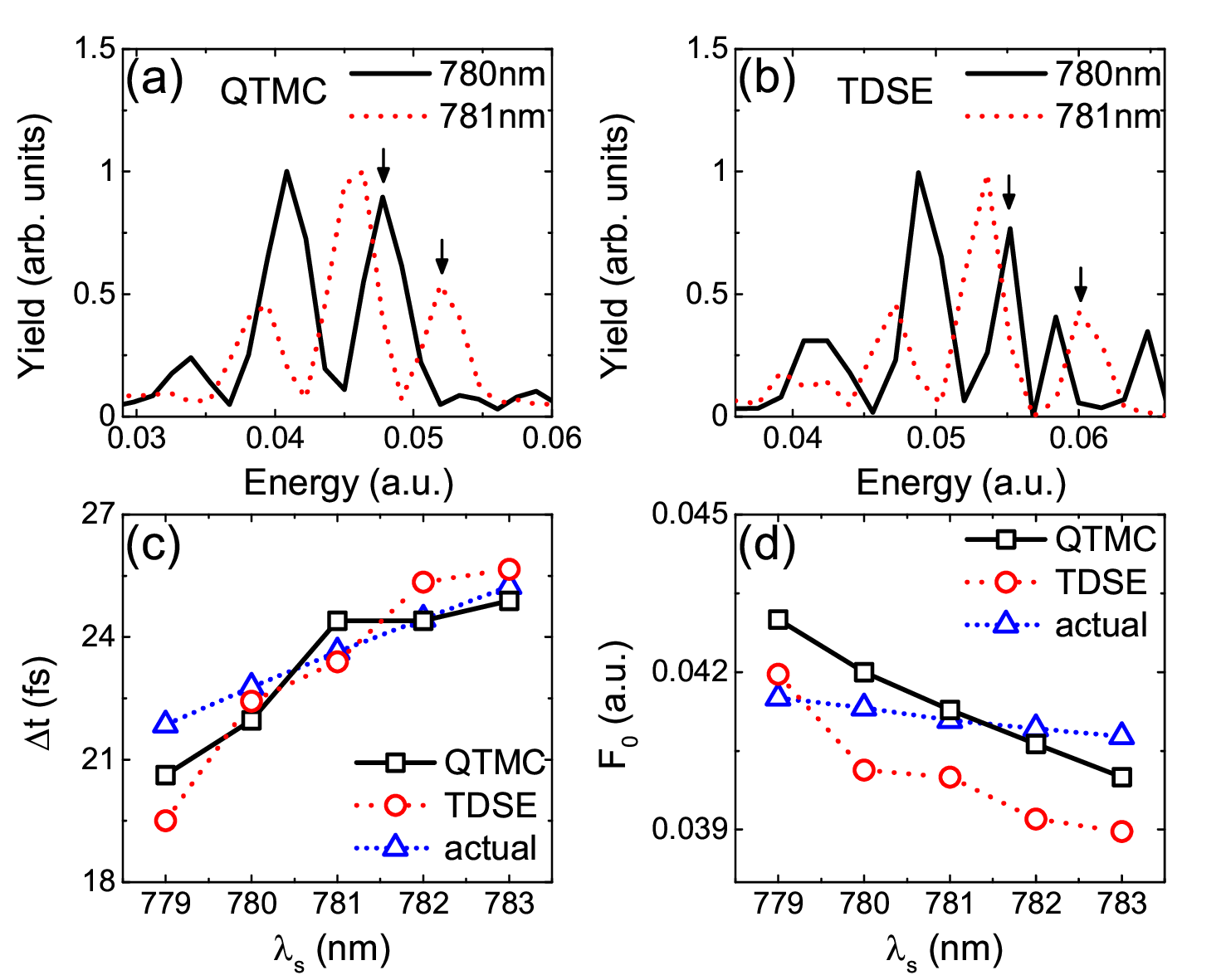}
\caption{Energy spectra for shaped pulses simulated by QTMC (a) and TDSE
(b), and the corresponding extracted time delay $\Delta t$ (c) and peak
electric field strength $F_0$ (d) of the shaped pulse at different $\protect%
\lambda_{s}$. The actual values of $\Delta t$ and $F_0$ are also presented
for comparison. The arrows in panels (a) and (b) are used to indicate the
sub-ring used in Fig.~\protect\ref{fig4}.}
\label{fig3}
\end{figure}

Therefore, we only have to read the fringe spacing in the PMD to obtain the
time delay. Furthermore, if we know the exact value of $m$ in Eq. (\ref%
{fringe}), we can determine $U_{p}$ and thus determine the peak electric
field $F_{0}$ of the shaped pulse. This can be achieved with help of the
jet-like structure of the PMD. The jet-like structure has been widely
discussed previously and can be explained by either multiphoton ionization
or sub-cycle electron wave-packet interference \cite{Bai2006,Huismans2013}.
The nodes of the jet-like structure on the ATI ring is closely related to
the number of photons the electron absorbs. The case of 10 nodes in Fig.~\ref%
{fig1}(b) corresponds to 11-photon channel. We can deduce a
guessed $U_{p}$ according to the energy conservation: $n\omega
_{0}-I_{p}-U_{p}=E$, which describes that a bounded electron absorbs $n$
photons to overcome the binding potential $I_{p}$ plus the pondermotive
potential $U_{p}$ and eventually becomes a free electron with energy $E$.
Then we substitute the guessed $U_{p}$ into Eq.~(\ref{fringe}) to make
correction to it by ensuring that $m$ is an integer. Eventually, both the
peak electric field and the time delay of the two sub-pulse are obtained for
specific $\lambda _{s}$. If varying $\lambda _{s}$, the shape of the
tailored laser field will change accordingly, resulting in the shift of the
interference fringes and the change of the fringe spacing. This can be seen
clearly in the angle-integrated energy spectra simulated by both QTMC [Fig.~%
\ref{fig3}(a)] and TDSE [Fig.~\ref{fig3}(b)] methods, in which the peaks
shift to the right and the spacing between them decreases when $\lambda _{s}$
increases. Applying the temporal Young's double-slit interferometer, $\Delta
t$ and $F_{0}$ for different $\lambda _{s}$ are extracted from the simulated
photoelectron spectra, as shown in Fig.~\ref{fig3}(c) and Fig.~\ref{fig3}%
(d). For comparison, we also present the actual values read directly from
the shaped pulse. The retrieved results for both QTMC and TDSE are found to
be in good agreement with the actual values, which proves the accuracy of
the interferometer. Even better, it needs only to obtain the slope of the $%
\lambda _{s}$ dependence with some raw data points in real experiments,
since the time delay and peak electric field vary with $\lambda _{s}$ almost
linearly. For example, the slope of the time delay vs $\lambda _{s}$ in Fig.~%
\ref{fig3}(c) is approximately 1097 as/nm, then for the shortest step of adjusting $%
\lambda _{s}$ that can be achieved, i.e., 0.2 nm, the corresponding step of
the time delay is 220 as.

\begin{figure}[tb]
\centering
\includegraphics[scale=0.44]{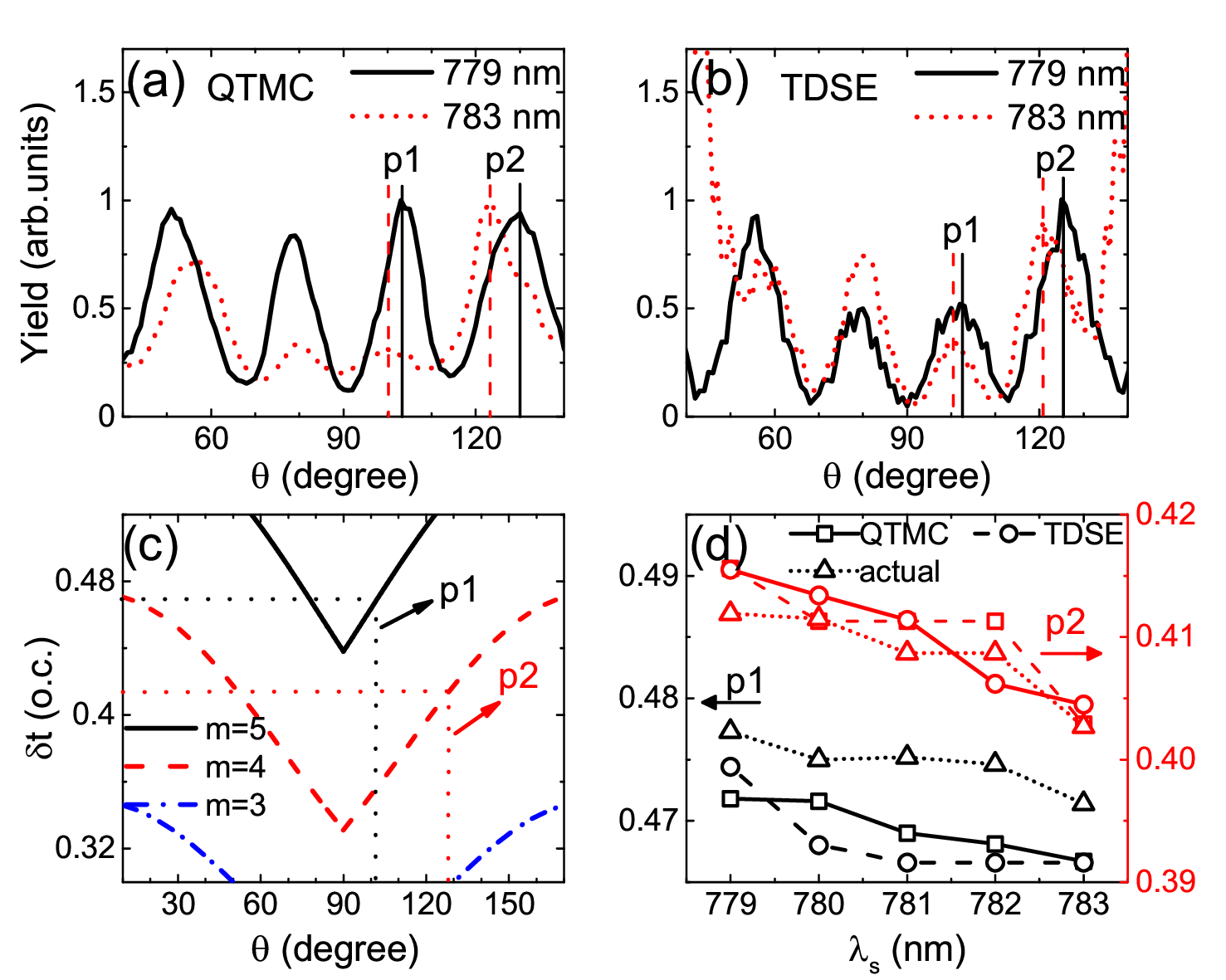}
\caption{Angular distribution corresponding to a specific sub-ring at
different $\protect\lambda_{s}$ simulated by QTMC (a) and TDSE (b). The
positions for peaks $p1$ and $p2$ are indicated by vertical lines. Panel (c)
shows the curves obtained from Eq. (\ref{dt}), and illustrates how
the ionization time difference $\protect\delta t$ between the two
interfering electron trajectories contributing to the two peaks (labelled as
$p1$ and $p2$) in panels (a) and (b) can be extracted from the curves. (d)
The extracted $\protect\lambda_{s}$-dependent $\protect\delta t$
corresponding to peaks $p1$ and $p2$. The actual values of $\protect\delta t$
obtained through statistics of the trajectories in the QTMC simulations are
also presented for comparison. The sub-rings used above are indicated by
arrows in Figs.~\ref{fig3}(a) and \ref{fig3}(b).}
\label{fig4}
\end{figure}

After accurate characterization of the electric-field waveform of the shaped
pulse, the precise control of electronic dynamics becomes possible.
Actually, the $\lambda _{s}$-dependent sub-ring structure in the PMD has
demonstrated the capacity of the shaped pulse to control the dynamics of the
tunneling wave packets. However, since the sub-ring structure is the result
of intercycle interference, the precision of this kind of manipulation of
the electron motion is limited to half the optical cycle (1.3 fs for an 800
nm laser), although the precision of modulating the time delay between the
two sub-pulses can be achieved on the attosecond scale. In fact, along with
the ability to control the time delay of the shaped pulse, we can also
precisely modulate its peak electric field strength concurrently [see Fig.
\ref{fig3}(d)]. Here, we will exploit the latter capability to control the
electron sub-cycle dynamics with attosecond precision. For this purpose, let
us turn to the angular distribution featured with a series of peaks
corresponding to the jet-like structure in the PMD, which is sensitive to
the peak electric field. Examining the jet-like structure for a specific
sub-ring, we find that the peaks in the photoelectron angular distribution
simulated by both TDSE and QTMC shrink towards 90 degrees (perpendicular to
the laser polarization) with increasing $\lambda _{s}$, as shown in Figs.~\ref{fig4}(a) and \ref{fig4}(b). Since the jet-like structure is the result
of intracycle interference, its evolution with $\lambda _{s}$ indicates the
possibility of manipulating the electron dynamics with sub-cycle precision.
This sub-cycle dynamic is characterised by $\delta t$, the difference in
ionization time for the two interfering electron trajectories born within an
optical cycle as sketched in Fig. \ref{fig2}(a2), which can be extracted from the angular distribution by phase
analysis based on Eq. (\ref{act}). The peak at a specific angle $\theta $ in
the angular distribution occurs when the phase difference between the two
trajectories experiencing intracycle interference reaches $2m\pi $, which
gives the following relationship (see supplement 1 for details)
\begin{equation}
(E+I_{p}+U_{p})\delta t-\frac{\sqrt{2E}F_{0}}{\omega _{0}^{2}}\left\vert
\cos \theta \right\vert \sin (\frac{\omega _{0}}{2}\delta t)-\frac{U_{p}}{%
\omega _{0}}\sin (\omega _{0}\delta t)=2m\pi.  \label{dt}
\end{equation}

As long as the photoelectron energy $E$ and the peak electric field strength
$F_{0}$ are specified, the above relationship between $\delta t$ and $\theta
$ for certain $m$ is determined. Here, this relationship is established
using the mean values of $E$ and $F_{0}$ at different $\lambda _{s}$. The
curves given by Eq.~(\ref{dt}) are shown in Fig.~\ref{fig4}(c), from which $%
\delta t$ can be directly extracted. According to the phase analysis, if we
count the number of peaks in the angular distribution starting from 90
degrees, the first peak corresponds to the largest $m$ that Eq. (\ref{dt})
allows, the second peak corresponds to $m-1$, and so on. Therefore, $\delta
t $ for peak $p1$ in Fig.~\ref{fig4}(a) should be extracted from the curve
of $m=5$, and $\delta t$ for peak $p2$ should be read from the curve of $m=4
$, as illustrated in Fig.~\ref{fig4}(c). In Fig.~\ref{fig4}(d), we present
the $\delta t$ extracted from spectra simulated by QTMC and TDSE at
different $\lambda _{s}$, which exhibits a monotonously decreasing
dependence on $\lambda _{s}$. We also provide the actual values of $\delta t$
obtained through statistics of the trajectories contributing to the peaks in
the QTMC simulations. The good agreement between the extracted $\delta t$
for both QTMC and TDSE with the actual values supports the validity of the
extraction method. Note that there is a greater discrepancy between the
extracted and actual values for peak $p1$ compared to peak $p2$. This is due to
the fact that the additional rescattering trajectories not considered in Eq. (\ref{dt}) are involved in the formation of peak $p1$
\cite{li2014}, which will affect the extracted results. The results in Fig.~\ref{fig4}(d) clearly demonstrate that
the sub-cycle electronic dynamics, characterized by $\delta t$, can be
manipulated with attosecond precision by regulating the shaped laser field.
The slope of the regulation curve for $\delta t$ extracted from peak $p1$
and peak $p2$ is approximately 4 and 7 as/nm, respectively.

\section{Conclusions}

In summary, we proposed a temporal Young's double-slit interferometer to
characterize the shaped laser field with unprecedented precision. By
reversing the phase of the frequency spectra from specific wavelength $%
\lambda_{s}$, a conventional single femtosecond pulse is split into a pair
of sub-pulses in time domain after Fourier transformation, whose shape can
be precisely controlled by adjusting $\lambda_{s}$. Based on the principle
of Young's double-slit interference in time domain, in which the two
sub-pulses are analogous to the double-slit, the peak electric field and the
time delay between them can be precisely retrieved from the interference
pattern in PMD resulted from the interaction of shaped pulse with atoms.
With this capability, we show that the sub-cycle dynamics of electron can be
controlled with shaped pulse and the precision of control is in attosecond
scale. The above scheme is proved to be feasible by both QTMC and TDSE
simulations. It is worthy to note that the shaped pulse consisting of two
sub-pulses can also be regarded as the construction of a pump-probe scheme. Our
work demonstrates that the time delay between the pump and probe pulses can
be manipulated in attosecond precision. In addition, the proposed scheme can
also be extended to circularly polarized laser field to combine with the
technology of attoclock, which may be capable of probing and controlling
electronic dynamics with even higher precision.

\section*{Funding.}

This work was supported by the National Key Research and Development Program
(No.~2019YFA0307700), the National Natural Science Foundation of China
(No.~12274273, No.~12204314, No.~92261201, No.~12274179), the Innovation Program for Quantum Science and Technology (No.~2021ZD0302101) and
the Natural and Science Foundation of Top Talent of SZTU(No.~GDRC202202).
\section*{Disclosure.}

The authors declare no conflicts of interest.

\section*{Data availability.}

Data underlying the results presented in this paper are not publicly
available at this time but may be obtained from the authors upon reasonable
request.

\section*{Supplemental document.}

See Supplement 1 for supporting content.

%\section*{Supplemental document.} See Supplement 1 for supporting content.
%%%%%%%%%%%%%%%%%%%%%%% References %%%%%%%%%%%%%%%%%%%%%%%%%

%Add references with BibTeX or manually.
%\cite{Zhang:14,OSA,FORSTER2007,Dean2006,testthesis,Yelin:03,Masajada:13,codeexample}
%%%%%%%%%% If using BibTeX:
%\bibliography{ref}

%%%%%%%%%% If preparing manually:

\end{document}